\documentclass[pra,twocolumn,showpacs,preprintnumbers,amsmath,amssymb,article]{revtex4-1}
\pdfoutput=1
\usepackage{graphicx}
\usepackage{float}
\usepackage{verbatim}
\usepackage{dcolumn}
\usepackage{bm}
\usepackage{dsfont}

\begin{document}


\newcommand{\beq}{\begin{equation}}
\newcommand{\eeq}{\end{equation}}
\newcommand{\beqa}{\begin{eqnarray}}
\newcommand{\eeqa}{\end{eqnarray}}
\newcommand{\lf}{\hfil \break \break}
\newcommand{\ahat}{\hat{a}}
\newcommand{\adag}{\hat{a}^{\dagger}}
\newcommand{\adagg}{\hat{a}_g^{\dagger}}
\newcommand{\bhat}{\hat{b}}
\newcommand{\bdag}{\hat{b}^{\dagger}}
\newcommand{\bdagg}{\hat{b}_g^{\dagger}}
\newcommand{\chat}{\hat{c}}
\newcommand{\cdag}{\hat{c}^{\dagger}}
\newcommand{\dhat}{\hat{d}}
\newcommand{\nhat}{\hat{n}}
\newcommand{\ndag}{\hat{n}^{\dagger}}
\newcommand{\den}{\hat{\rho}}
\newcommand{\phihat}{\hat{\phi}}
\newcommand{\Ahat}{\hat{A}}
\newcommand{\Adag}{\hat{A}^{\dagger}}
\newcommand{\Bhat}{\hat{B}}
\newcommand{\Bdag}{\hat{B}^{\dagger}}
\newcommand{\Chat}{\hat{C}}
\newcommand{\Dhat}{\hat{D}}
\newcommand{\Ehat}{\hat{E}}
\newcommand{\Lhat}{\hat{L}}
\newcommand{\Nhat}{\hat{N}}
\newcommand{\Ohat}{\hat{O}}
\newcommand{\Odag}{\hat{O}^{\dagger}}
\newcommand{\Shat}{\hat{S}}
\newcommand{\Uhat}{\hat{U}}
\newcommand{\Udag}{\date{\today}
\hat{U}^{\dagger}}
\newcommand{\Xhat}{\hat{X}}
\newcommand{\Zhat}{\hat{Z}}
\newcommand{\Xdag}{\hat{X}^{\dagger}}
\newcommand{\Ydag}{\hat{Y}^{\dagger}}
\newcommand{\Zdag}{\hat{Z}^{\dagger}}
\newcommand{\Ham}{\hat{H}}
\newcommand{\bis}{{\prime \prime}}
\newcommand{\tris}{{\prime \prime \prime}}
\newcommand{\ket}[1]{\mbox{$|#1\rangle$}}
\newcommand{\bra}[1]{\mbox{$\langle#1|$}}
\newcommand{\ketbra}[2]{\mbox{$|#1\rangle \langle#2|$}}
\newcommand{\braket}[2]{\mbox{$\langle#1|#2\rangle$}}
\newcommand{\bracket}[3]{\mbox{$\langle#1|#2|#3\rangle$}}
\newcommand{\mat}[1]{\overline{\overline{#1}}}
\newcommand{\dotp}{\mbox{\boldmath $\cdot$}}
\newcommand{\tp}{\otimes}
\newcommand{\op}[2]{\mbox{$|#1\rangle\langle#2|$}}
\newcommand{\hak}[1]{\left[ #1 \right]}
\newcommand{\vin}[1]{\langle #1 \rangle}
\newcommand{\abs}[1]{\left| #1 \right|}
\newcommand{\tes}[1]{\left( #1 \right)}
\newcommand{\braces}[1]{\left\{ #1 \right\}}

\hyphenation{Teich}

\title{SU(2) uncertainty limits}
\author{Saroosh Shabbir}
\email[e-mail:]{saroosh@kth.se}
\affiliation{Department of Applied Physics, Royal Institute of Technology (KTH)\\
AlbaNova University Center, SE - 106 91 Stockholm, Sweden}
\author{Gunnar Bj\"{o}rk}
\affiliation{Department of Applied Physics, Royal Institute of Technology (KTH)\\
AlbaNova University Center, SE - 106 91 Stockholm, Sweden}

\date{\today}

\begin{abstract}
Although progress has been made recently in defining nontrivial uncertainty limits for the SU(2) group, a description of the intermediate states bound by these limits remains lacking. In this paper we enumerate possible uncertainty relations for the SU(2) group that involve all three observables and that are, moreover, invariant under SU(2) transformations. We demonstrate that these relations however, even taken as a group, do not provide sharp, saturable bounds. To find sharp bounds, we systematically calculate the variance of the SU(2) operators for all pure states belonging to the $N=2$ and $N=3$ polarisation excitation manifold (corresponding to spin 1 and spin 3/2). Lastly, and perhaps counter to expectation, we note that even pure states can reach the maximum uncertainty limit.
\end{abstract}


\maketitle

\section{Introduction}

The algebra of SU(2) is ubiquitous in Physics, describing, among other phenomena, the angular momentum characteristics of atomic systems. Formally equivalent to the angular momentum operators, the Stokes parameters describe the polarisation state of light. In classical optics, the degree of polarisation has been defined as the length of the Stokes vector. This quantitative measure of polarisation has been found to be insufficient in the quantum domain; states which appear unpolarised have been shown to possess polarisation structure via higher (than dipolar) order correlation measurements \cite{Klyshko,Usachev,Tsegaye}. Thus, it is essential to consider quantum fluctuations of the Stokes operators to arrive at an operationally adequate description of polarisation in the quantum domain. Moreover, quantum fluctuations of angular momentum operators, and hence their associated uncertainty relations, also play a crucial role in metrology, where they define the ultimate limit to the resolution of interferometric measurements \cite{Yurke,Wineland,Rivas}. In this paper, we study the second-order statistics to arrive at nontrivial limits for the uncertainty relation for the Stokes observables and detail the states that reside within these limits.\\

In some sense the uncertainty relations in quantum mechanics embody the departure from the classical world. They describe the impossibility of simultaneous sharp preparations of incompatible observables as embodied in the Robertson-Schr\"{o}dinger (R-S) uncertainty relations \cite{Robertson,Schrodinger}. Since the SU(2) group has three mutually incompatible generators, one can write three different uncertainty relations for different pairs of observables using the R-S inequalities. However, a more natural uncertainty relation involving all three generators cannot be arrived at by using only their commutation relations. Moreover, in contrast to the canonical uncertainty relations involving position and momentum observables, $\Delta x \Delta p_x \geq \hbar/2 $, the uncertainty relations for the SU(2) group give state-dependent bounds which can lead to trivial results. Below we present a framework to address these issues.

Non-canonical uncertainty relations have been studied in various ways. One approach has considered the states saturating the R-S uncertainty relation \cite{Trifonov,Puri,Kinani} named as ``intelligent states'' \cite{Trifonov}. However, although the class of intelligent states seems to have applications in interferometry with sensitivity below the shot noise level \cite{Hillery,Brif}, they don't seem to have attracted the experimentalists' fancy. Another approach has been to form weighted uncertainty relations \cite{Maccone}. These sometimes provide sharper bounds than the Robertson-Schr\"{o}dinger relations, but at the expense of weighing the operators unequally. An experimental justification for doing so is presently lacking. In \cite{Chen} uncertainty limits were derived for an arbitrary number of non-commuting observables, but the limits were state dependent, just as in the intelligent state approach. Pati \textit{et al.} \cite{Pati} have used another approach and derived the uncertainty limits for a joint measurement of many identical copies of a certain state. They have shown that such a joint measurement of $N$ identical systems will have an uncertainty a factor of $N^{-1/2}$ smaller than if each of the $N$ systems were measured individually and these $N$ measurement values were used to estimate the value of the measured observable.

Our approach in this work is to some extent similar to W\"{u}nsche's who has derived higher-order uncertainty relations for a variety of algebras from invariants \cite{Wunsche}. Specifically, much of the algebra needed to derive our relations is found in \cite{Rivas}, where the authors looked at the ultimate measurement precision of angular momentum operators, which apart from a factor of 1/2 are identical to the Stokes operators. The alternative method of quantifying uncertainties via entropic uncertainty relations has also been studied extensively \cite{Bialynicki,Deutsch,Maassen}, however here we restrict ourselves to non-entropic measures.\\

\section{Stokes operators and their uncertainty relations}
\label{Sec: Stokes operators}
For the monochromatic field of light in two orthogonal modes, the Stokes operators \cite{Stokes,Collett} can be succinctly represented as
\beq
\hat{S}_{\mu} = (\adag_R \hspace{2mm} \adag_L)\  \sigma_{\mu} \begin{pmatrix}
           \ahat_R
        \vspace{1.5mm}\\
           \ahat_L
           \end{pmatrix},
\eeq
where $\ahat_R$ and  $\ahat_L$ are the annhilation operators of right (R) and left handed (L) circular polarisation modes. The Greek letter $\mu$ runs from 0 to 3 with $\sigma_0=$ \(\mathds{1}\), and ${\sigma_k}$, $k \in \{1,2,3\}$, are the Pauli matrices. The Casimir operator $\Shat_0$ then defines the total number of photons, and the Stokes vector $\langle{\bf \Shat} \rangle =\langle \Shat_1,\Shat_2,\Shat_3\rangle$ indicates the polarisation in horizontal/vertical, diagonal/anti-diagonal and left-/right-circular modes respectively. (The Schwinger boson representation of the angular momentum operators are smaller than the Stokes operators by a factor of 1/2, but otherwise identical.) Using these operators, the Stokes operator in a arbitrary (normalized) direction $\mathbf{n}$ on the Poincar\'{e} sphere can be written $\Shat_\mathbf{n} = \mathbf{\Shat} \cdot \mathbf{n}$.

${\Shat_k}$ satisfy the commutation relations of the SU(2) algebra: $[\Shat_k,\Shat_l]= i \epsilon_{klm} \Shat_m$, where $\epsilon_{klm}$ is the Levi-Civita fully antisymmetric tensor. Since the commutation relation is state-dependent, so is the uncertainty limit for the Stokes (or angular momentum) operators
\beq
\Delta S_k \Delta S_l \geq \frac{1}{2}\left |\epsilon_{klm} \langle \Shat_m \rangle \right|,
\label{Eq: UR}
\eeq
where $\Delta S_k = (\langle \Shat_k^2 \rangle - \langle \Shat_k\rangle^2 )^{1/2}$. Thus, the uncertainty limit depends on the state, as mentioned above.\\

Complete second-order statistics of the polarisation observables can be extracted from the $3 \times 3$ covariance matrix $\mathbf{\Gamma}$ \cite{Feller,Rivas,covariance tensor}, where
\beq \label{covariance matrix}
\Gamma_{kl}= \langle \Shat_k \Shat_l + \Shat_l \Shat_k \rangle/2 - \langle \Shat_k \rangle \langle \Shat_l \rangle.
\eeq
It's utility stems from its simple connection to measurements and, furthermore, it's Hermitian by construction. Since it is a second-rank tensor, we can readily define three invariants; the determinant, the sum of the principle minors and the trace. Expressed in terms of the (real and non-negative) eigenvalues $\lambda_k$ of $\mathbf{\Gamma}$, $k \in \ \{1,2,3\}$, these can be used to form state-dependent uncertainty relations, viz:
\beq
0 \leq  \lambda_1 \lambda_2 \lambda_3  \leq \langle \Shat_0^3 (\Shat_0 + 2)^3 \rangle /27,
\label{Eq: First uncertainty}
\eeq
\beq \label{Eq: Second uncertainty}
 \Shat_0^2 \leq \lambda_1 \lambda_2 + \lambda_2 \lambda_3 + \lambda_3 \lambda_1  \leq \langle \Shat_0^2(\Shat_0 + 2)^2 \rangle/3,
\eeq
\beq
\label{Eq: sum variance}
2 \langle\hat{S}_0 \rangle \leq  \lambda_1 + \lambda_2 + \lambda_3 \leq  \langle \hat{S}_0(\hat{S}_0+2) \rangle.
\eeq
The eigenvalues $\lambda_k$ are the principal variances in the above equations. If we use the corresponding orthonormal eigenvectors $\mathbf{\Lambda_k}$ of the covariance matrix on the Poincar\'{e} sphere, then the variance of the Stokes operator $\Shat_\mathbf{n}$ can be written as
\beq
(\Delta \Shat_\mathbf{n})^2 = \sum_{k=1}^3 \left ( \mathbf{\Lambda_k} \cdot \mathbf{n} \right )^2 \  \lambda_k.
\eeq
This implies that, e.g., the smallest (largest) of the three eigenvalues for a certain state will define the smallest (largest) Stokes operator variance under any polarization rotation of the state. However, while the relations \eqref{Eq: First uncertainty}-\eqref{Eq: sum variance} are state dependent, all three relations are invariant under any polarisation transformation. We will come back to this important point below.

Eq. \eqref{Eq: sum variance} is the more restrictive among the three uncertainty relations enumerated above. For example, the lower limit of \eqref{Eq: First uncertainty} follows from the non-negativity of the eigenvalues, and this limit is reached for all SU(2) coherent states that are eigenstates of one of the Stokes operators and thus have zero variance in that observable. Given the constraint $\lambda_1 + \lambda_2 + \lambda_3 \leq  \langle \hat{S}_0(\hat{S}_0+2) \rangle$ from \eqref{Eq: sum variance} and the fact that the eigenvalues are real and non-negative, choosing the eigenvalues to be equal, i.e., $\langle \hat{S}_0(\hat{S}_0+2) \rangle/3$, will maximize their product. The upper limit of \eqref{Eq: First uncertainty} follows, and this limit is reached by pure states such as the polarisation NOON-states $(\ket{N,0} + \ket{0,N})/\sqrt{2}$ which is perhaps counter-intuitive since one would expect the uncertainty to be maximum for only mixed states. To derive the lower limit of \eqref{Eq: Second uncertainty} one notes that all three bilinear uncertainty terms are non-negative, so to minimize the sum, the best one can do with a term is to make it vanish. If one of the principal variances, say $\lambda_1$, is zero, two of the terms vanish and do not contribute to the sum. However, this implies that the state in question is an eigenstate of $\Shat_1$, which in turn implies that $\lambda_2 = \lambda_3$. To get the lower limit for \eqref{Eq: Second uncertainty} we should now try to simultaneously minimise $\lambda_2$ and $\lambda_3$. We therefore look at the lower limit of \eqref{Eq: sum variance} from which we find that the smallest permissible values for $\lambda_2 = \lambda_3$ is $S_0$, which inserted in \eqref{Eq: Second uncertainty} defines its lower limit. To find the upper limit of \eqref{Eq: Second uncertainty} we see that the maximum under the sum constraint \eqref{Eq: sum variance} is reached when $\lambda_1=\lambda_2=\lambda_3$ and using the upper limit \eqref{Eq: sum variance} we see that the maximum hence is reached when all eigenvalues equal $S_0(S_0 + 2)/3$. Inserting this into \eqref{Eq: Second uncertainty} leads to its upper limit. In the following we utilise \eqref{Eq: sum variance} and in particular the fact that the trace operation is basis independent, \textit{so the sum of the Stokes operator variances is equal to the sum of the principal variances and is constant, invariant of any polarization transformation of the state}.  We use this invariance to demonstrate the uncertainty structure of all pure states for a given excitation manifold. However, equations \eqref{Eq: First uncertainty}-\eqref{Eq: sum variance} are not particularly ``sharp'' in denoting what principal variance triplets $\left( \lambda_1, \lambda_2, \lambda_3 \right)$ are permissible. As an example, the triplet $(0.75, 0.75, 2.5)$ satisfies all three equations, but no pure, two-photon state has this particular variance triplet.

\section{Poincar\'{e} sphere and the Majorana representation}
The Poincar\'{e} sphere affords an elegant pictorial representation of the polarisation state of light. Each of the Stokes operators define the polarisation of the state along the three coordinate axes of the Poincar\'{e} Sphere. The polarisation characteristics of a single photon are represented by a point on, or within, the surface of the Poincar\'{e} sphere. For $N>1$, any pure, two-mode state can be represented on the Poincar\'{e} sphere via its Majorana representation \cite{Majorana,Zimba} which maps the $N$-photon state as $N$ points on the surface of the sphere. We know that any general, pure, two-mode $N$-photon state can be expressed as
\begin{align}
\label{Majorana rep}
\ket{\Psi_N} &= \sum_{n=0}^N c_N \ket{n,N-n} \\ \notag
&= \frac{1}{\sqrt{\mathcal{N}}} \prod_{n=1}^N \left( \cos(\theta_n/2) \adag_R + e^{i \phi_n} \sin(\theta_n/2) \adag_L \right) \ket{0,0},
\end{align}
where $\mathcal{N}$ is the normalisation factor \cite{Hofmann,Shabbir}. Each of the factors in the above equation can be represented as a point on the Poincar\'{e} sphere with coordinates ($\sin(\theta_n)\cos(\phi_n), \sin(\theta_n)\sin(\phi_n), \cos(\phi_n)$). For example, the Majorana representation of an $N$-photon, SU(2) coherent state, $\ket{\Psi_{\textrm{SU(2)}}} = (N!)^{1/2}(\adag_R)^N \ket{0,0} $, is a collection of $N$ points stacked on top of each other at the North pole of the Poincar\'{e} sphere, in line with the intuitive description of SU(2) coherent states as exhibiting the most classical behaviour with their Stokes vector pointing in one particular direction.

At the opposite end, SU(2) maximally unpolarized, pure states have vanishing Stokes vector and isotropic variance (for $N >3$). That is to say their polarisation, up to second-order, points nowhere \cite{Zimba,stars}. The Majorana representation of these states is comprised of points spread as symmetrically as possible over the surface of the Poincar\'{e} sphere. For certain excitations greater than 3, these points form the vertices of the Platonic solids. Somewhere in-between these two extremes is the $N$-photon NOON state, $\ket{\Psi_{\textrm{NOON}}}= (2N!)^{1/2}((\adag_R)^N + (\adag_L)^N )\ket{0,0}$ which can be represented by $N$ equidistant points along the equator. It is because of this configuration that such NOON states have the highest sensitivity to small rotations about $\Shat_3$ \cite{stars}, thus underscoring their metrological importance. Similar to the biphotons $\ket{1,1}$ generated in spontaneous parametric down conversion, the NOON states manifest hidden polarisation \cite{Klyshko} for $N \geq 2$. All of these have vanishing Stokes vector but do indeed show polarisation structure with higher-order polarisation correlation measurements such as the variance.\\

\section{$N=2$ orbits and variance}
In terms of the Stokes operators any general, linear polarisation transformation can, e.g., be written as:
\beq
\label{polsarisation trans}
\hat{U}_{Pol} = \exp(i \alpha \hat{S}_3) \exp(i \beta \hat{S}_2) \exp(i \gamma \hat{S}_3)
\eeq
Such transformations rotate the input state around $\hat{S}_3$ by an angle $2\gamma$, followed by a rotation around $\hat{S}_2$ by $2\beta$ followed by a final rotation again around $\hat{S}_3$ by $2\alpha$. From this two conclusions follow: First, an SU(2) transformation \textit{rigidly} rotates the configuration of the points represented by \eqref{Majorana rep}, resulting in a different state on the same SU(2) orbit as the original state. To state explicitly, an \textit{orbit} is the locus of SU(2) transformations for a particular state. Described another way, an orbit is a set of all states that are mutually convertible via SU(2) transformations. Second, different orbits are parametrised by different \textit{relative} orientations of the Majorana points with respect to each other on the Poincar\'{e} sphere. For $N=2$, there is only one relevant parameter governing the orientation of the two points with respect to each other - the angle $\theta$ subtended by the two points at the centre of the sphere. The most general representation of the $N=2$ orbit generating state can thus be given by fixing one of its Majorana points at the North pole and constraining the other to move on the Greenwich-Meridian. This allows us to write the orbit generating state for the $N=2$ manifold as:
\beq
\label{N2 orbit gen state}
\ket{\Psi_2} = \frac{1}{\sqrt{\mathcal{N}}} \adag_R \left( \cos(\theta/2) \adag_R + \sin(\theta/2) \adag_L \right) \ket{0,0},
\eeq
where $0 \leq \theta \leq \pi$.
\begin{figure}[ht]
\includegraphics[scale=0.2]{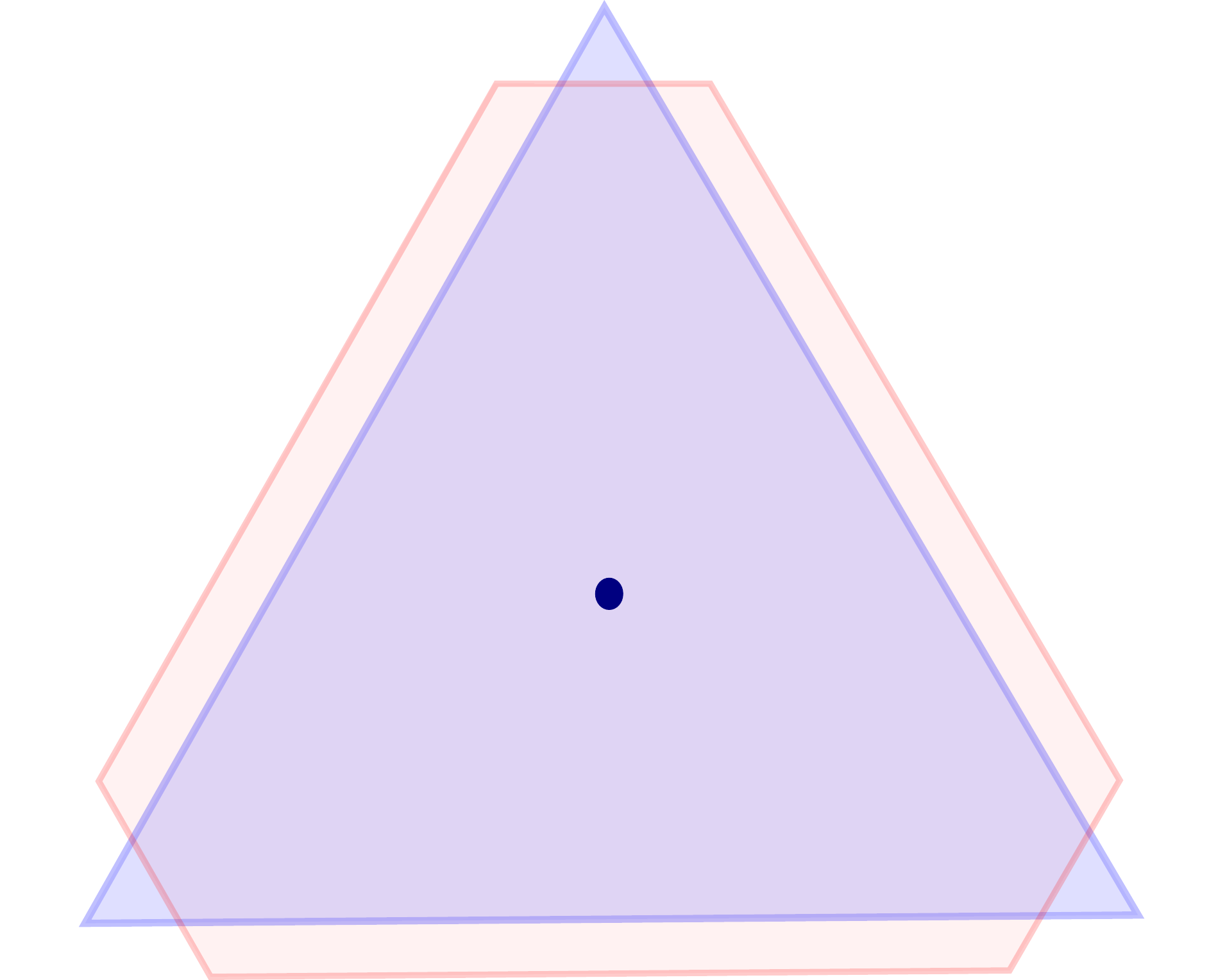}
\caption{Convex hull of the allowed variances. The eigenvalues of the covariance matrix form the vertices of the polygons. In case of threefold degeneracy, the point in the middle is obtained. A triangle is formed in the case of twofold degeneracy via cyclic permutation of the eigenvalues. If all three eigenvalues are distinct, one obtains an irregular hexagon. For $N\geq3$, the figure also demonstrates the situation of overlapping orbits (for more information see text).}
\label{Fig: orbits}
\end{figure}

Consequently, states with different $\theta$ lie on different SU(2) orbits. By definition, states on the same SU(2) orbits have the same properties for invariants such as the Stokes operator principal variance sum \eqref{Eq: sum variance}. As mentioned in Sec. \ref{Sec: Stokes operators}, this sum, being the trace of a tensor, is always equal to the sum of the variances for any state. Hence, $(\Delta S_1)^2 + (\Delta S_2)^2 + (\Delta S_3)^2$ is a constant on every orbit.

We know that the $N$-photon SU(2) coherent states and the $N$-photon NOON states saturate the lower and upper limits of Eq. \eqref{Eq: sum variance}, respectively. (Choosing $\theta=\pi$ in equation \eqref{N2 orbit gen state} defines the state $\ket{1,1}$, which can be transformed into the 2-photon NOON state by an SU(2) rotation. This is particular only to the $N=2$ excitation manifold.) To study the intermediate states, we calculate the covariance matrix from the orbit generating state as a function of the orbit generating parameter $\theta$. The eigenvalues of the covariance matrix ($\lambda_1, \lambda_2, \lambda_3$) give the extrema of the Stokes operator variances. These can be physically permuted by applying an SU(2) transformation that rotate the state around a particular Stokes-operator axis by $\pm \pi/2$. In variance space formed by the axes $(\Delta S_1)^2$, $(\Delta S_2)^2$, and $(\Delta S_3)^2$, using permutations of eigenvalues as the vertices, the so-constructed polygon is the convex hull of the allowed variances for a given orbit (Fig. \ref{Fig: orbits}). Thus, all the points inside the polygon including the border are reachable from any other point on the polygon via an SU(2) transformation. 

\begin{figure}[h]
\includegraphics[scale=0.43]{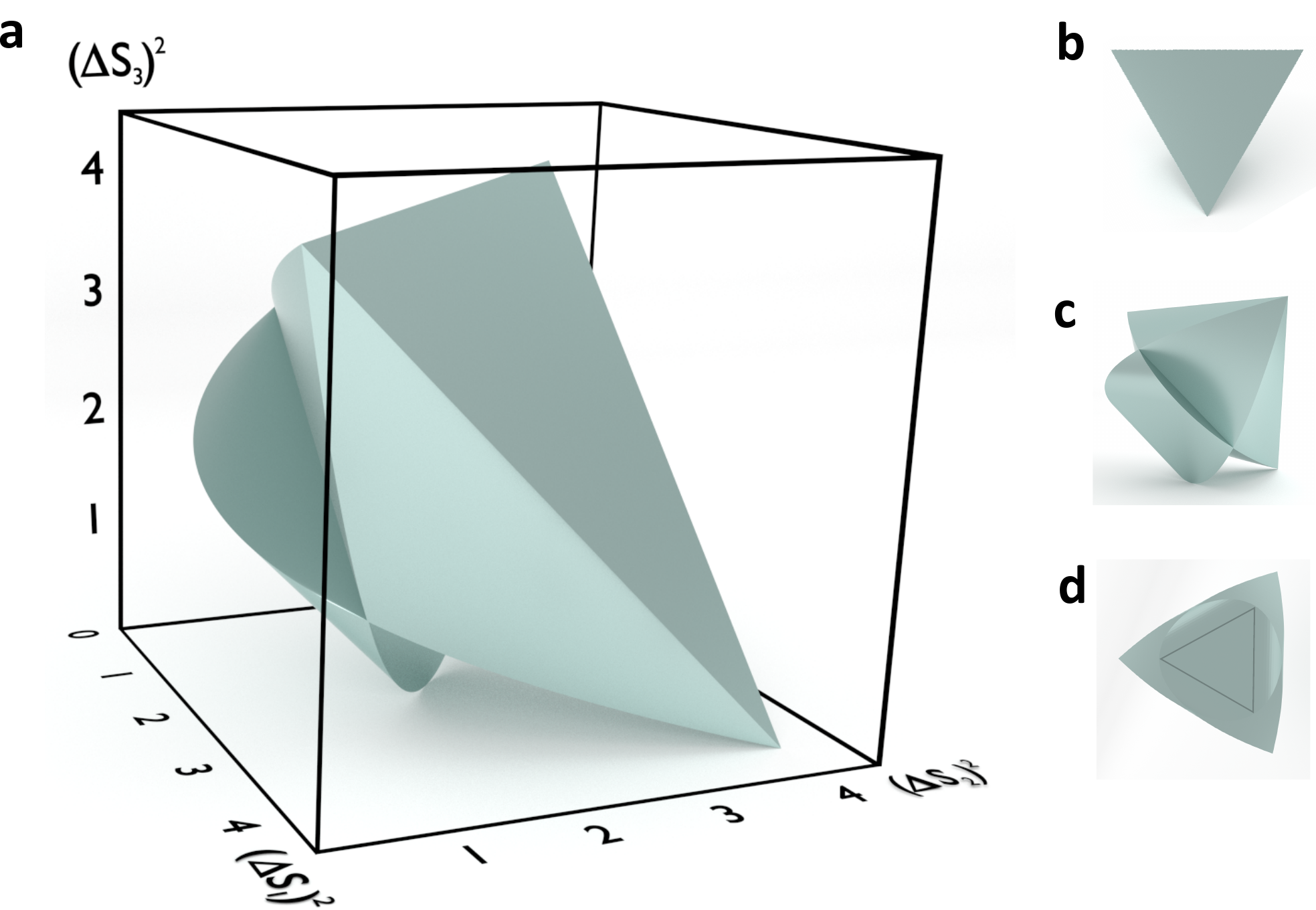}
\caption{(a) Permissible uncertainty values for the $N=2$ excitation manifold. The volume is formed by stacking the obtained polygons in Fig. \ref{Fig: orbits}. (b) Top view with surface normal in the (1,1,1) direction, corresponds to the orbit for the NOON state. (c) Side view. (d) Bottom view with surface normal again in the $(1,1,1)$ direction. The drawn triangle shows the orbit for the SU(2) coherent state.}
\label{Fig: N2}
\end{figure}

If the eigenvalues are threefold degenerate, $\lambda_1 = \lambda_2 = \lambda_3$, one obtains a point. In other words the states on such an orbit have isotropic variance over the Poincar\'{e} sphere that will not change under any SU(2) rotation. Doubly degenerate eigenvalues lead to a triangle in variance space. Due to the fact that the sum of the three variances is a constant, the triangle's surface normal will point in the direction $(1,1,1)$ in the variance space, thus keeping the variance sum constant on the triangle surface. In the case of no eigenvalue degeneracy, the orbit defines a triangle with chopped corners (in general, an irregular hexagon) in variance space, again oriented in the $(1,1,1)$-direction. Finally, to obtain the whole volume of allowed variances, polygons for different orbit parameter values $\theta$ in Fig. \ref{Fig: orbits} are stacked on top of each other, and the hull of the stacked polygons is drawn. For the $N=2$ excitation manifold the stacking is relatively easy to do, as we know that the sum variance, that determines the distance between each polygon's center and the origin in Fig. \ref{Fig: N2}, is monotonically increasing with an increase of the orbit generating parameter $\theta$ in the interval $0 \leq \theta \leq \pi$.

Fig. \ref{Fig: N2} shows the shape of the volume in which all permissible triplets of variances $((\Delta S_1)^2,(\Delta S_2)^2,(\Delta S_3)^2)$ must lie for $N=2$. Any cross-section normal to the variance space-diagonal is a polygon circumscribing the allowed variances for a specific orbit. The polygon closest to the origin is a triangle, with the variance sum 4 and the vertices (2,2,0), (2,0,2) and (0,2,2). As expected, it represents the SU(2) coherent state $\ket{2,0}$ and its SU(2) orbit. The polygon furthest from the origin is also a triangle, with variance sum 8 and the vertices at (4,4,0), (4,0,4) and (0,4,4). This triangle represents the state $\ket{1,1}$ and its orbit. On each orbit one can quite obviously find a state which has $(\Delta S_1)^2 = (\Delta S_2)^2 = (\Delta S_3)^2$. However, for $N=2$ the parameter space is too small to allow for complete isotropy, that is, an orbit with invariant Stokes variances.

\section{$N=3$ orbits and variance}
The Majorana representation of 3 points corresponds to a triangle on the surface of the Poincar\'{e} sphere, and as a result the orbit generating state for $N=3$ is a function of three parameters. Following the same idea as for the $N=2$ case, we fix one point on the North pole ($\theta_1=\phi_1=0)$. The second point is constrained to move on the Greenwich-Meridian ($\phi_2=0)$ and the third point is nominally allowed all possible $\theta$ and $\phi$ configurations. Accordingly, the orbit generating state for the $N=3$ manifold is given as follows:
\begin{eqnarray}
\label{N3 orbit gen state}
\ket{\Psi_3} & = & \frac{1}{\sqrt{\mathcal{N}}} \adag_R \left( \cos (\theta_2/2 ) \adag_R + \sin(\theta_2/2 ) \adag_L \right) \nonumber \\
&& \times \left( \cos(\theta_3/2 ) \adag_R + e^{i \phi_3}  \sin(\theta_3/2 ) \adag_L \right) \ket{0,0},
\end{eqnarray}
where one can use the restriction $\theta_2 \leq \theta_1$ to eliminate some of the degeneracy this parameterisation leads to. It is also possible to restrict $\phi \leq \pi$ since point configurations obeying mirror symmetry define identical uncertainty limits although they don't belong to the same orbit. The stacking of the polygons is a bit trickier in this case, as different combinations of the orbit generating states parameter may result in overlapping orbits (that is, orbits with the same variance sum). An explicit example will be given below.

\begin{figure}[h]
\includegraphics[scale=0.43]{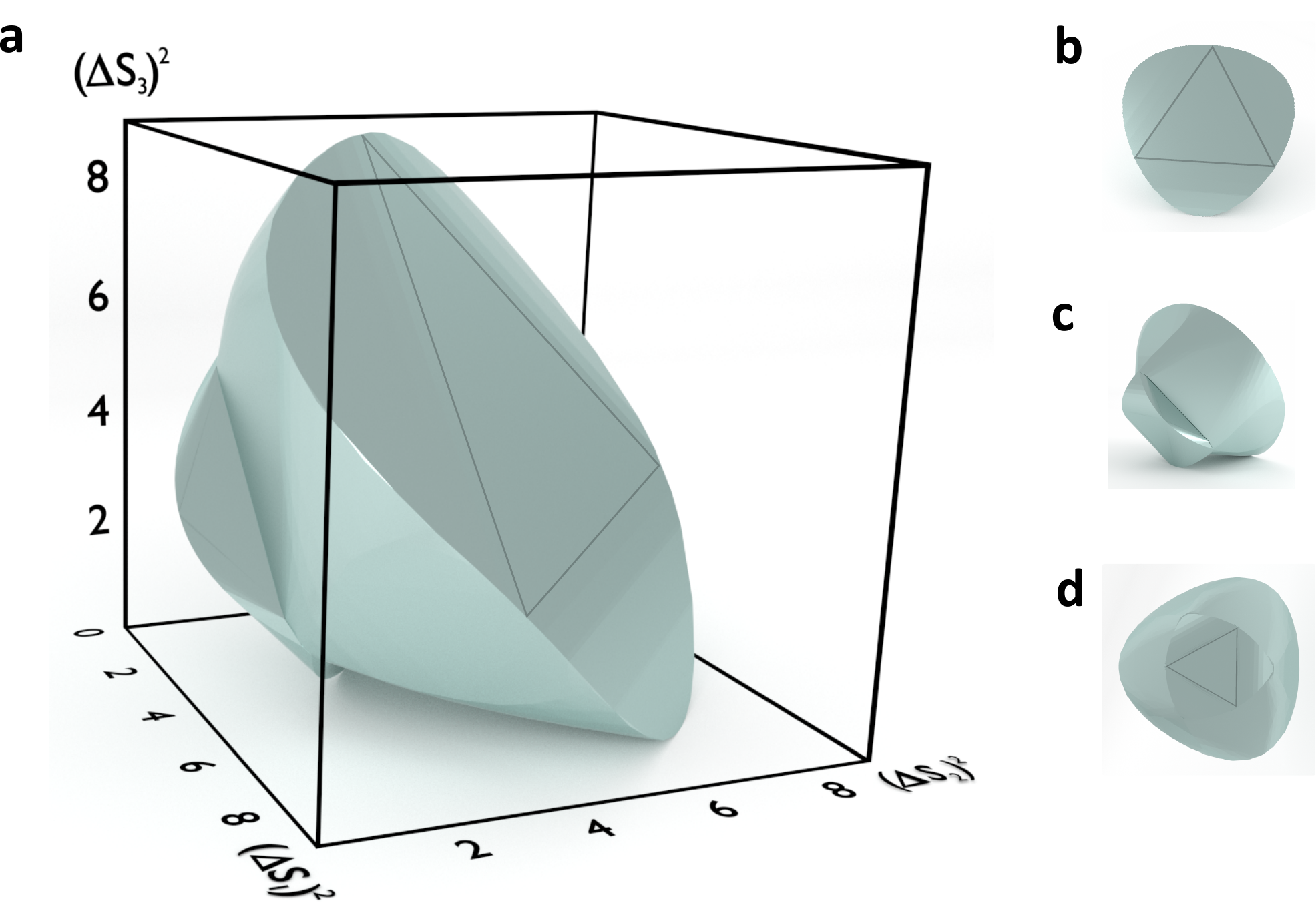}
\caption{(a) Permissible uncertainty volume for the $N=3$ manifold. (b) Top view with surface normal in the (1,1,1) direction. The drawn triangle shows the orbit for the NOON state. (c) Side view. (d) Bottom view with surface normal again in the $(1,1,1)$ direction. The orbit for the SU(2) coherent state is shown.}
\label{Fig: N3}
\end{figure}
We see that the permissible uncertainty values for $N=3$ has a similar structure as for the $N=2$ case (Fig. \ref{Fig: N2}). However, as just mentioned, one of the primary distinctions is that in contrast to the $N=2$ case, the cross-section normal to the space diagonal is composed of two or more overlapping orbits. As a consequence, different orbits may have the same variance sum as illustrated in Fig. \ref{Fig: orbits} where, e.g., one orbit could have its permissable variances bounded by a triangle, while a different orbit with the same sum variance may be bounded by a irregular hexagon. Thus, equal sum variance is not a sufficient condition for two states to be on the same orbit and thus be mutually transformable via an SU(2) transformation.

An interesting class of states
\beq
\ket{\eta_N} = \eta \ket{N,0} + \sqrt{1-\eta^2} \ket{0,N},
\eeq
where
\beq
\eta=\sqrt{\frac{1}{2} \left(1 \pm \sqrt{(N-1)/N}\right)},
\eeq
emerges in the $N \geq 3$ excitation manifolds. These states have the Stokes vector $(0, 0, N(2 \eta^2 - 1))$, but have isotropic variance
\beq \Delta \Shat_\mathbf{n}^2 = N,
\eeq 
for any SU(2) transformation $\hat{U}_{Pol}$ where $\Shat_\mathbf{n}=\hat{U}_{Pol}^\dagger \Shat_1 \hat{U}_{Pol}$. Such states are thus uniform (variance) states \cite{Zimba} and their orbit is represented by a point in the variance coordinate system. However, as their Stokes vector does not vanish, they are not ``anti-coherent'' \cite{Zimba}, i.e., they are not uniform states with vanishing mean polarisation. These states may be interesting for polarimetric applications since their variance does not change with the rotation of the Stokes vector. In contrast, maximally unpolarised, pure states \cite{stars} have vanishing Stokes vector \textit{and} isotropic variance. However, states unpolarised to second-order can only be found for $N = 4$ and $N \geq 6$ \cite{stars}.

For $N=3$ the uniform state can be written as \beq \ket{\eta_3}=(0.30291, 0, 0, 0.95302), \eeq expressed in the basis $\left \{\ket{0,3}, \ket{1,2}, \ket{2,1}, \ket{3,0}  \right \}$, with a non-vanishing Stokes vector $(0,0,2.4495)$. If the state is rotated such that one Majorana point lies at the North pole, and another along the Greenwich meridian, then the state's Majorana points are defined by the angles $\theta_1 = 0$, $\theta_2 = \theta_3 \approx 107.5^\circ$, $\phi_2=0$, and $\phi_3 \approx 115.47^\circ$.

This state has the same variance sum ($\Delta \Shat_1^2+\Delta \Shat_2^2+\Delta \Shat_3^2=9$) as the state $(0, 0.5704, 0.7914, 0.2199)$ that lies on another orbit. The latter state's principal variances are $\lambda_1=1.1637$, $\lambda_2=1.8990$, and $\lambda_3=5.9373$. Thus, the state's associated variance orbit is an irregular hexagon. These two states demonstrate the concept of different orbits' overlapping sum variance, as illustrated in Fig. \ref{Fig: orbits}.

\section{Conclusions}

In conclusion, we have illustrated the permissible Stokes operator variances for all pure states for the $N=2$ and $N=3$ manifold. The obtained figures are surprisingly involved and show that uncertainty relations such as \eqref{Eq: UR} or \eqref{Eq: First uncertainty}-\eqref{Eq: sum variance} have limited value. Since the Stokes operators obey the SU(2) algebra, all such operators, for instance the angular momentum operators, will be restricted to the same, or proportionally scaled, uncertainty volumes. The method is extendible to higher excitation manifolds, but gets progressively difficult. As as example, for the $N=4$ excitation manifold, one can define the orbit generating state in the same way as for the $N=2$ and $N=3$ cases. The obit generating state will then be a function of 5 parameters and all distinct orbits will correspond to distinct Majorana point quadrilaterals on the Poincar\'{e} sphere. The difficulty however lies in defining angular limits that result in distinct orbits. An alternative approach is to generate a point cloud sampled sufficiently densely over all orbits and all polarisation transformations (which adds another three parameters). Such an analytically simplistic strategy comes at the expense of significantly increased computation time.

We have highlighted a simple method to check whether states lie on the same SU(2) orbits and are connected via linear polarisation transformations in terms of the equivalence of their Majorana representations. This provides a first check for identifying realisable quantum experiments \cite{Krenn} (excluding post-selection) in the domain of quantum optics. In this context, consequently, the sum uncertainty relation equation \eqref{Eq: sum variance} provides useful experimental information; given different values for the uncertainty sum for the initial and required states, one can be sure that (excluding all non-unitary processes such as post-selection etc.) there can no be experimental realisation that creates the required state from the initial.

Lastly, pertaining to mixed states we find that although their uncertainty is also bounded by the limits in \eqref{Eq: sum variance}, they do not always lie inside the shown uncertainty volumes.

\begin{acknowledgments}
G.B. acknowledges fruitful discussions with Profs. L.~L.~S\'anchez-Soto, and A.~B.~Klimov. This work was supported by the Swedish Research Council (VR) through grant 621-2014-5410 and through its support of the Linn\ae us Excellence Center ADOPT.\\
After submission, the recent work by Dammeier \textit{et al.} \citep{Dammeier} which has a few results similar to this paper was brought to our attention.
\end{acknowledgments}

\end{document}